\documentclass[prl,amsmath,showpacs, amsfonts,twocolumn,preprintnumbers, superscriptaddress]{revtex4}
\usepackage{amsfonts}
\usepackage{color,graphicx}
\usepackage{bm}
\usepackage{latexsym, amsfonts, amssymb}

\newcommand{\Ce}{CeCoIn$_5$}

\newcommand{\hc}{H_{c_2}}
\newcommand{\hp}{H_{Pauli}(0)}

\begin{document}
\title{Evidence for the FFLO state in \Ce\  from penetration depth measurements}

\author{C. Martin}
\author{C. C. Agosta}
\email{cagosta@clarku.edu} 
\affiliation{Department of Physics, Clark University, Worcester, MA,  01610}
\author{ S. W. Tozer}
\author{ H. A. Radovan}
\author{E. C. Palm}
\author{T. P. Murphy}
\affiliation{The National High Magnetic Field Laboratory,
1800 E. Paul Dirac Dr., Tallahassee, FLÊ 32310}
\author{J. L. Sarrao}
\affiliation{Los Alamos National Laboratory, Los Alamos, New Mexico 87545}
\date{\today}

\begin{abstract}
We report penetration depth and resistivity measurements on the heavy fermion superconductor \Ce\ using a self resonant tank circuit based on a tunnel diode oscillator. For
magnetic fields applied near parallel to the \emph{ab}-planes and temperatures below 250 mK, two phase transitions were found. The lower field transition, within the superconducting state, is of a second
order and we identify it as the transition from the ordinary vortex state to the Fulde, Ferrell, Larkin, Ovchinnikov (FFLO) state. The higher field transition marks the change from the FFLO to the normal state. This higher field transition, $\hc$, is of first order up to 900 mK, the highest temperature measured. Our normal state resistivity measurements at temperatures between 100 and 900 mK suggest that the FFLO state is related to the change of the quasi-particle interaction strength, $F^a_0$. Our critical field data for different orientations and temperatures is in good agreement with recent specific heat, magnetization, and resistivity results.
\end{abstract}


\maketitle
In 1964, Fulde and Ferrell \cite{FF} and Larkin and Ovchinnikov
\cite{LOV} predicted that in a purely Pauli limited superconductor, the
magnetic field acting on the Cooper pair's spin can induce pairs with
nonzero total momentum and, consequently, a spatially
modulated order parameter. We have made penetration depth measurements suggesting that this so called FFLO state exists in the heavy fermion superconductor \Ce.
 
The FFLO state can lead to an enhancement of the critical field up to 2.5 times the Pauli
paramagnetic limit \cite{Shim}. Gruenberg \emph{et al}\cite{GG} showed in 1966 that
the FFLO state may exist even in the presence of orbital
effects. In the past few years, there has been
 a growing interest in the theoretical study of this more realistic case, when the
 critical field is determined by both paramagnetic and orbital
 effects \cite{Buzdin1, Shim2, Houzet1, Houzet2, Adachi, Manalo02}, and special
 interest in anisotropic (Quasi-2D) \cite{Buzdin1, Shim2, Houzet1, Adachi, Manalo02}
 and d-wave superconductors \cite{Shim2, Adachi, Manalo02}, which are properties found in \Ce.  
The theoretical literature suggests that organics and heavy-fermion superconductors are very promising
candidates for observing the FFLO state. In heavy-fermion materials the
presence of this state may be even more favorable \cite{HBurk, Buzdin1}.
Their very low Fermi velocity (large effective mass) decreases the
orbital-magnetic field coupling, often ensuring that the
system is Pauli limited. However, the first unambiguous experimental
evidence for the formation of the FFLO state has been reported
only very recently \cite{Henri} from specific heat and magnetization measurements on \Ce. More recently, a paper was published by Bianchi et al., that supports this claim \cite{Bianchi3}.
Most of the previous experimental results on \Ce\ ($T_c=2.3K$) \cite{Petrovic1} are in good
agreement with the theoretical criteria for observing the FFLO
state. It has been shown that the orbital pair breaking effect has
to be small or completely absent, as measured by the  Maki parameter $\alpha=\sqrt{2}\frac{\hc^0}{\hp}$  \cite{Maki}, where $\hc^0$ is the orbital critical field \cite{Bianchi1}, to favor the FFLO state. This would translate into a value of $\alpha$$\geq 1.8$ according to Ref. \onlinecite{GG}, although this calculation was partially based on the BCS theory and may not apply to non s-wave superconductors. This ratio, $\alpha$, has been calculated by a number of groups \cite{Bianchi1, Henri}, but the parameters for this ratio, particularly $\hp$, are difficult to measure. 
From magnetization studies of the critical
field \cite{Tim, Tayama}  it is clear that, when the magnetic
field is applied parallel to the \emph{ab}-planes ( the \emph{ab}-planes are perpendicular to the [001] direction) at low temperature, $\hc$ is a first order transition, indicating that \Ce\
is in the Pauli limit.  We have calculated the Pauli limit for \Ce\ by using a theory-independent method \cite{zuo, Agosta2}, that requires knowledge of Wilson's ratio \cite{mckenzie}. The basis of the calculation is that the condensation energy, $U_c$, is related to both the specific heat via the specific heat jump, and the binding energy of the Cooper pairs via, 
 \begin{equation}
  U_c  = \frac{ \mu_0}{2}\chi_eH^2_{Pauli},
 \end{equation}
 where  $\chi_e$ is the Pauli paramagnetic susceptibility. We propose that the jump in the measured specific heat can be integrated from the superconducting transition to zero temperature as an estimate of the condensation energy of the superconducting state, and using the measured susceptibility, $H_{Pauli}$ can be calculated. Although the specific heat leads to a good measure of the condensation energy,  the susceptibility measures the Pauli susceptibility plus unwanted terms such as the  Landau diamagnetism and inner core electrons. One way to isolate the Pauli susceptibility is to use the Sommerfeld constant $\gamma$ to estimate $\chi_e$ through the use of Wilson's ratio. The difficulty with this method is that the Land\'e g factor \cite{mckenzie} needs to be measured to find Wilson's ratio, and we are unaware of any direct measurements of g in \Ce. However,  Won et. al \cite{WonMaki} have recently published a critical field calculation, which they fit to critical field data with g as one of the few free parameters. In this paper they found g = 0.64 in the parallel orientation and 1.5 in the perpendicular orientation. Using data for $\gamma, \Delta C$, and $H_{c2}(0)$ ($H^0_{c2||} = 31.5$ T and  $H^0_{c2\bot} = 12.5$ T.)\cite{Henri, Petrovic1, Tayama}, we calculated the Pauli limit in \Ce\ to be 7.3 T  and 4.8 T  yielding $\alpha  = 6.1$ and 3.7 when B is parallel and perpendicular to the \emph{ab}-planes, respectively.  Higher values of $\hp$ can be justified if \Ce\ is a not a Fermi Liquid as is discussed below.  Lower values of $\hp$ are found if $\chi_e$ is used directly  from recent measurements, which results in unphysical calculated values. These Pauli limits show that with H perpendicular to the \emph{ab}-planes, where the critical field was found to be 5.0 T, $\hc$ is near the Pauli limit, but in the  orientation with H along the \emph{ab}-planes $\hc$ is above the Pauli limit, consistent with an FFLO state.  
 
The enhancement of the upper critical field in the FFLO state can be particularly substantial for a 2D superconductor \cite{Bulaevskii} and de Haas-van Alphen data on \Ce\ indicate a pronounced
quasi-two-dimensional Fermi surface \cite{Hall,Settai}. Specific heat data \cite{Movshovich}, and thermal conductivity measurements \cite{Izawa}, suggest the presence of nodes in the
superconducting gap, which is important to characterize the FFLO state \cite{HBurk, Buzdin1, Shim}.
Another requirement for the FFLO state is that the superconductor be in the \emph{clean} limit, $l \gg \xi $, where $l$ is the mean free path of the quasi particles and $\xi$ is the superconducting coherence length. With $l\geq$~810~\AA\ and $\xi\leq$~58 \AA\ based on Refs.~\onlinecite{Movshovich} and \onlinecite{Tayama}, \Ce\ meets this requirement.  
 
In this letter we present tunnel diode oscillator (TDO) measurements on \Ce\  to fields of 18 tesla, at different orientations and temperatures
between 60 and 900 mK. The TDO is a self resonant tank circuit where the sample is placed in the  coil with the \emph{ab} planes perpendicular to the ac magnetic field \cite{Coffey}. In this orientation, the penetration depth is measured parallel to the \emph{ab} planes. Crystal platelets of \Ce\ with approximate dimensions of 1.0 mm dia $\times$ 0.1 mm  thick
 and 0.3 mm dia $\times$ 0.1 mm  thick 
  were placed in a 1.33 mm and a 0.35 mm diameter coil, respectively. Details of the sample growth and characterization can be found in Ref. \onlinecite{Petrovic1}. A thermometer was placed on the rotating platform, close to the
sample, so that we were able to account for the dc power added by the
TDO circuit. The data reported in this letter comes from the large sample where the resonant frequency of the circuit at 80 mK was
 $\simeq $189 MHz. The smaller sample and coil at 1.2 
GHz yielded similar results.  For a TDO experiment,  the relative change
in the resonant frequency is proportional to the change in the
penetration depth ($\frac{\Delta f}{f_0}=\eta\frac{\Delta
\lambda}{\lambda_0}$). Obtaining absolute values of the penetration
requires a careful calculation of the constant $\eta$, which depends on
the coil and sample geometries and the demagnetization factor. We
did not calibrate the system for obtaining absolute values, but we
measured and subtracted the influence of the background by running
the system both with and without the sample, so that the change in
the frequency follows accurately the change in penetration
depth of only the sample. In the normal state, the
penetration is limited  by the skin depth ($\delta\propto\sqrt{\rho}$), and we were able to measure the
change in resistivity with magnetic field and temperature. 

\begin{figure}[h] \begin{center}
\includegraphics[keepaspectratio=1,width=8 cm]{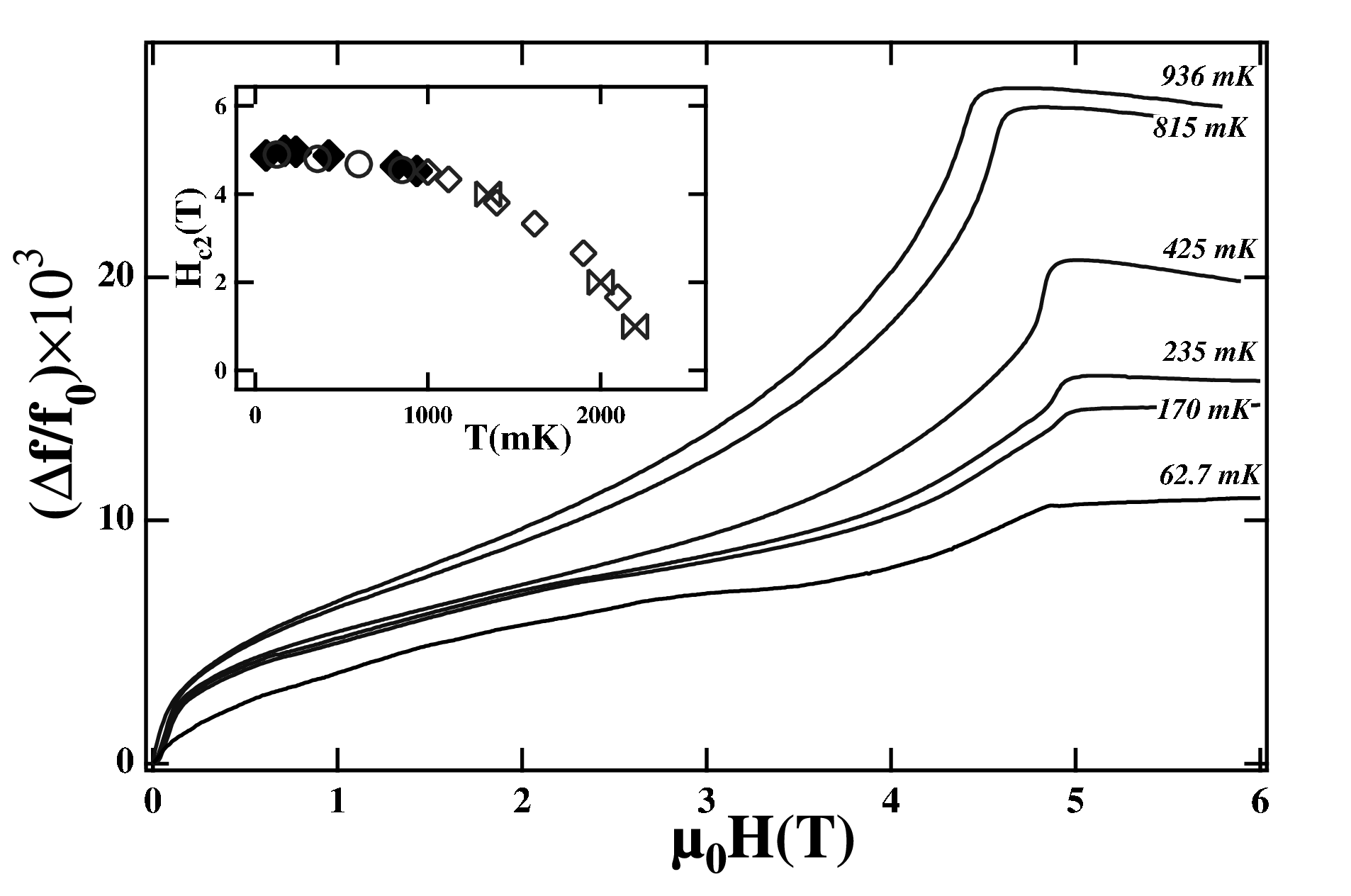}
\caption{\label{Perpen}The relative change in frequency with 
magnetic field at different temperatures for H perpendicular to the \emph{ab} 
planes. The two sweeps above 800 mK do not feature a clear 
jump in frequency at H$_{c2}$. The inset displays the $\hc$ vs temperature for this orientation. The filled diamonds represent our data, and the open diamonds, circles 
and bowties are magnetization data from Ref. \cite{Tayama}.}
\end{center} \end{figure}

Our phase diagram H-T for the field applied perpendicular to the \emph{ab} planes is in very good agreement with the magnetization data \cite{Tayama} as is shown in Fig.~\ref{Perpen}. Analysis of our data shows that this transition changes from first order to second order between 425 and 815 mK, where we no longer observed a sharp jump in the frequency of $\hc$, which is similar to other experiments \cite{Tayama, Bianchi1}. To the lowest temperature measured, we do not see evidence for the FFLO state in the perpendicular direction, although we note that the shape of the 62.7 mK  sweep is different than the higher temperature sweeps. We intend to study the low  temperature region more carefully in the future.

\begin{figure}[h] \begin{center}
\includegraphics[keepaspectratio=1,width=8 cm]{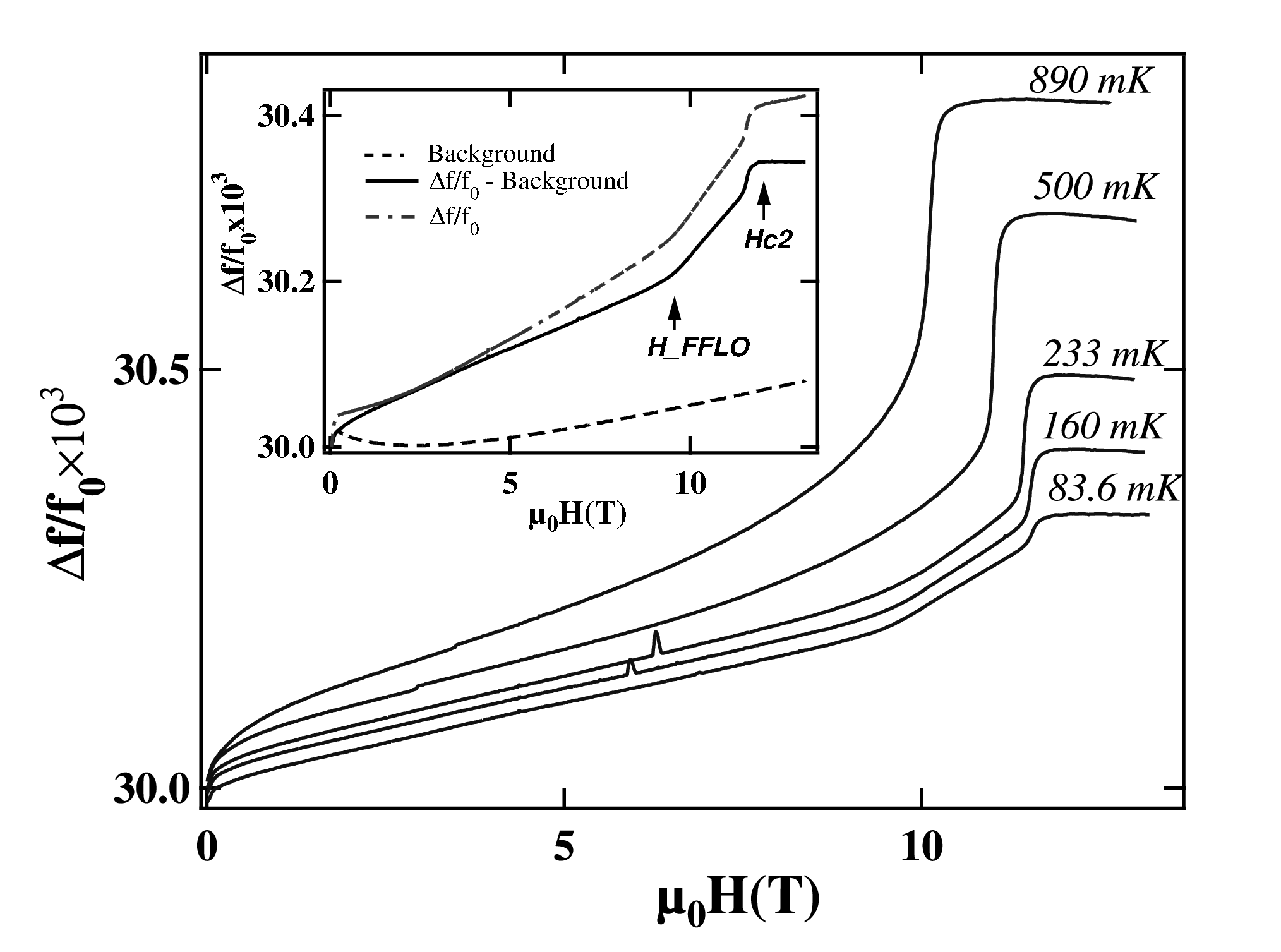}
\caption{\label{rawdata}The relative change in frequency with 
magnetic field at different temperatures for H // \emph{ab} planes. 
The inset shows the raw data at 83.6 mK, the background 
data, and the data with the background subtracted.}
\end{center} \end{figure}

Figure~\ref{rawdata} shows the relative change in frequency with magnetic
field when the field is applied parallel to the \emph{ab}-planes. The background and the
subtracted background data are also displayed.  As can be seen in Fig.~\ref{rawdata}, there is a lower field kink, or continuous transition reflected by the change of the slope of the
penetration depth versus field, followed by a sharp jump at higher field. We
assign the kink,  a second order transition, to the 
ordinary vortex state (VS)-FFLO transition and the upper transition, which is
first order, to the FFLO-Normal state transition. This is in very good agreement with
the recent observation of FFLO state in \Ce\ by specific heat
measurements \cite{Henri,Bianchi3}. Refs. \onlinecite{Adachi},  \onlinecite{Buzdin1} and \onlinecite{HBurk} 
have shown theoretically that, unlike originally predicted, the
transition at the lower critical field could be of second order, as seen in our
data.  Above T $\simeq$
250 mK, the second order transition is no longer present. The data shows only
the first order phase transition up to the dilution refrigerator limit of 900 mK, as seen in Fig.~\ref{rawdata}. This first order transition was first predicted by Maki for a type-II superconductor with a
large $\alpha$ parameter, below 0.56 T$_c$, albeit for the dirty limit \cite{Maki}. 

\begin{figure}[h] \begin{center}
\includegraphics[keepaspectratio=1,width=8 cm]{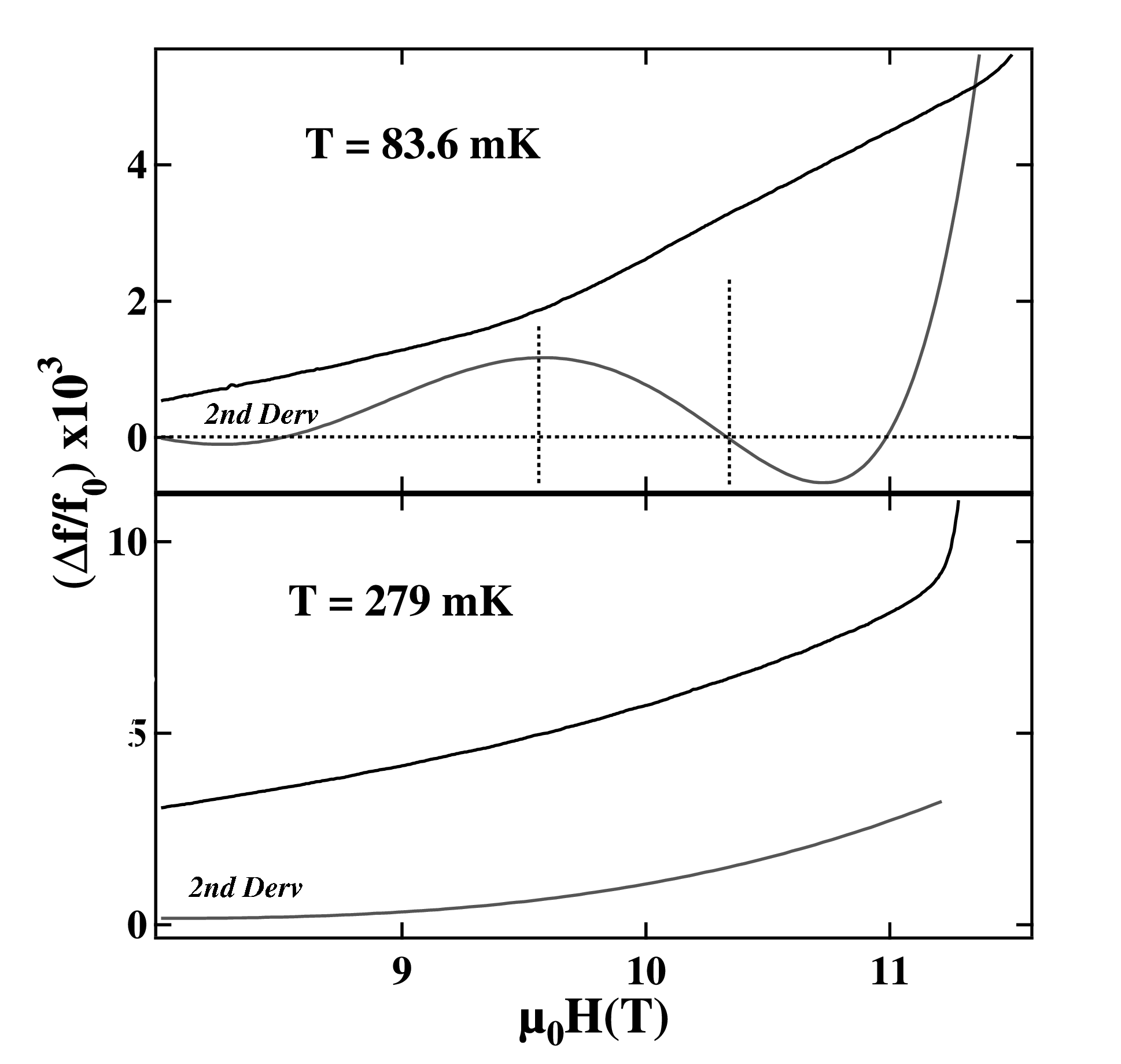}
\caption{\label{Derivatives}The second derivative of the penetration with respect to magnetic field  for two temperatures in the region around the FFLO transition. The positions of the 
local maximum and inflection point are marked 
in the upper region for clarity. At 279 mK, the second derivative is always positive.}
\end{center} \end{figure}

The distinguishing feature of the FFLO state is that the sign of the order parameter alternates spatially, either sinusoidally or more abruptly as the magnetic field approaches the FFLO-normal state phase line \cite{HBurk, Buzdin2}.
 The penetration depth, which is a function of the order parameter \cite{Tinkham}
($\lambda\propto\frac{1}{|\psi|}$  and $\propto\frac{1}{\sqrt{n_s}}$ ), is sensitive 
to the density of superconducting electrons, $n_s$.  If the order parameter oscillates and is no longer uniform, then the average density of superconducting electrons will be less and the penetration depth will increase.  Qualitatively this is the change in our data at the VS-FFLO transition. It is important to mention that the VS-FFLO transition occurs within the superconducting state where resistance measurements, of course, cannot see any signal change. 
 
\begin{figure}[h] \begin{center}
\includegraphics[keepaspectratio=1,width=8 cm]{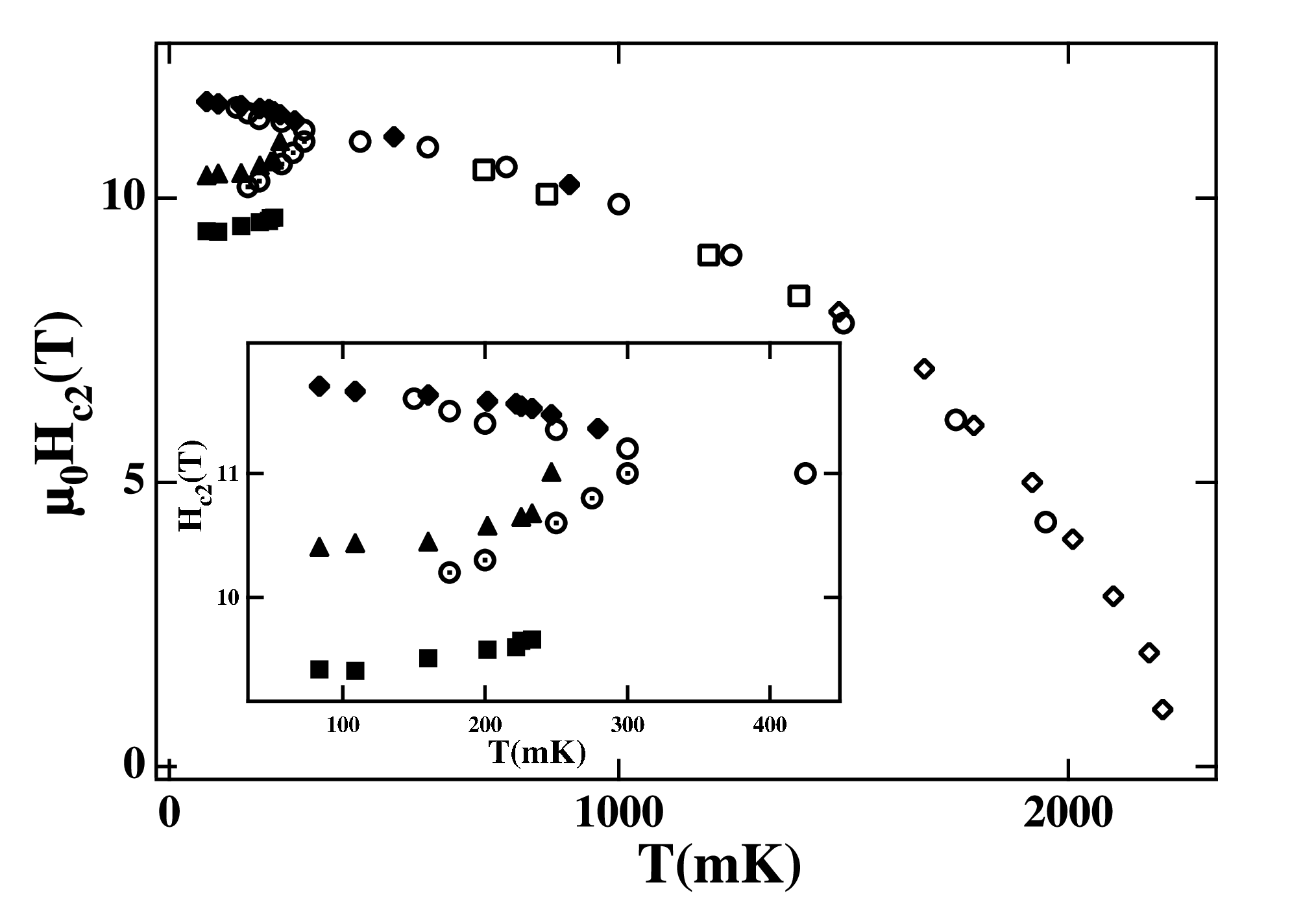}
\caption{\label{Phasedia}Critical field as a function of temperature 
for H // \emph{ab}. The filled  symbols represent our data: the 
upper critical field (diamonds), the FFLO transition from the 
point of inflection (equilateral triangles) and the FFLO from 
the position of the \emph{kink}, maximum of second order 
derivative (squares). Open symbols are data from other 
studies: specific heat from Ref. \cite{Henri} (circles and dotted circles) and 
magnetization \cite{Tayama} (diamonds) and \cite{Tim} (squares).}
\end{center} \end{figure}

\begin{figure}[h] \begin{center}
\includegraphics[keepaspectratio=1,width=8 cm]{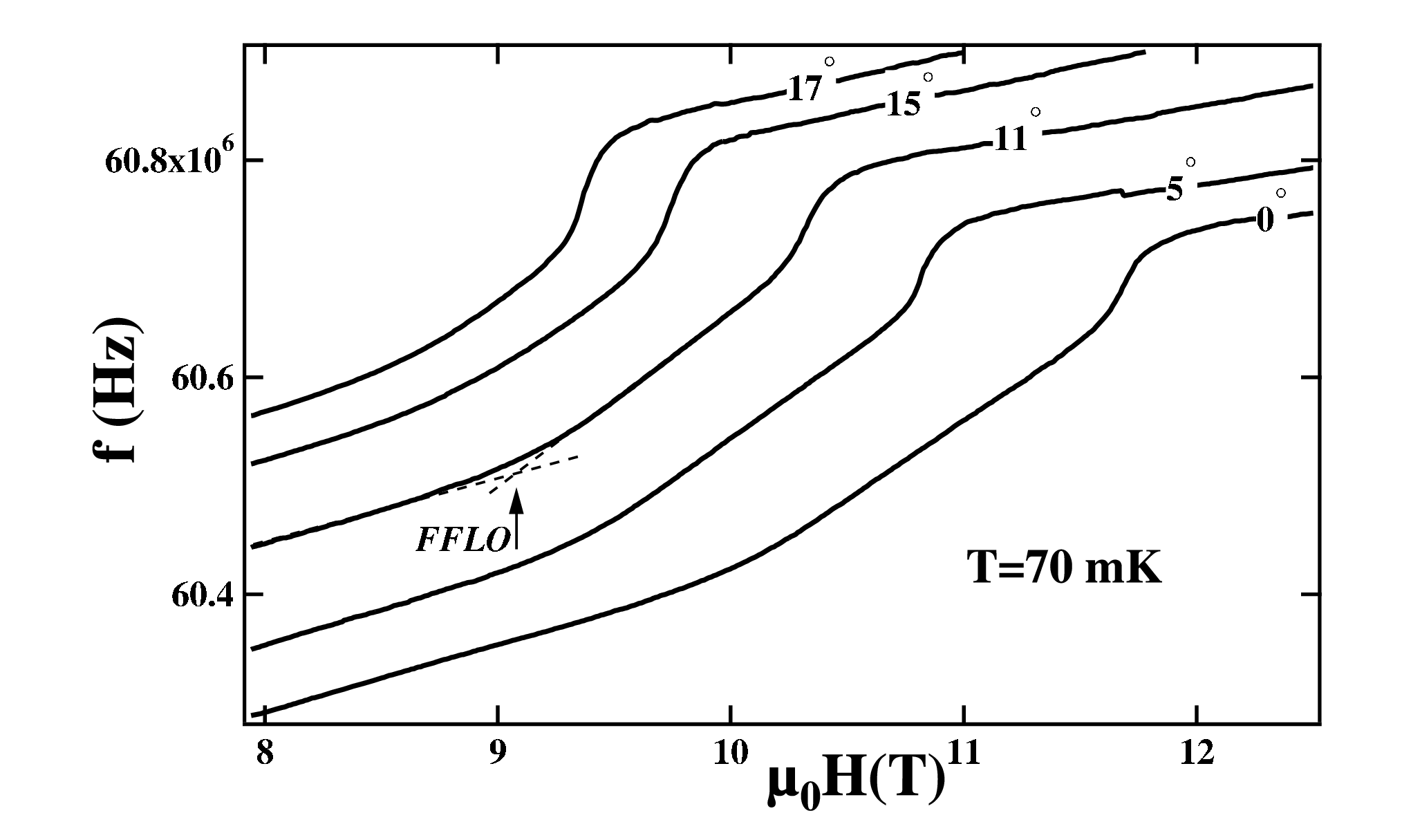}
\caption{\label{angledep}The penetration depth as a function of magnetic field applied at different angles with respect to the \emph{ab} planes. The FFLO transition goes away due to increasing orbital effects, however,  the superconducting to normal state transition is remains first order, as also was found in Ref. \cite{Tayama, Henri}.}
\end{center} \end{figure}

In Ref. \onlinecite{Henri} the FFLO state has been observed up to $T \sim 350 mK$ and
 B$_{FFLO}$ is almost 1 T higher than the values we obtain. One possible explanation for the difference in the position of the VS-FFLO transition may be the difference
in the quality of the samples, given the strong influence of impurities on the FFLO
state. On the other hand, this  difference is large, and it may be that the penetration depth and the specific heat measure different aspects of the VS-FFLO phase transition. To try and understand this discrepancy, we have calculated the second derivative of the penetration depth with respect to the field, $\lambda^{''}$, and show an example of $\lambda^{''}$ in Fig. \ref{Derivatives}. The first peak in $\lambda^{''}$ corresponds to the point of maximum curvature or the obvious kink in the penetration depth measurement near 9.5 T. The second easily identifiable feature in  $\lambda^{''}$ is where it crosses zero, signifying an inflection point. Although it is unclear exactly how $\lambda^{''}$ corresponds to what is physically happening in the FFLO phase transition, the zero of $\lambda^{''}$ corresponds to the end of the second order transition as measured by the TDO, and this point matches the transition as measured by specific heat data (see Fig. \ref{Phasedia}) \cite{Henri}. Given that  $\lambda^{''}$ is monotonic and has no zero crossings when the angle is greater than $15^o$ (see Fig.~\ref{angledep}) or the temperature is greater than 250 mK, and that the change in the slope of the penetration depth is consistent with an oscillating order parameter, this data provides clear evidence of an FFLO state in \Ce.  

\begin{figure}[h] \begin{center}
\includegraphics[keepaspectratio=1,width=8 cm]{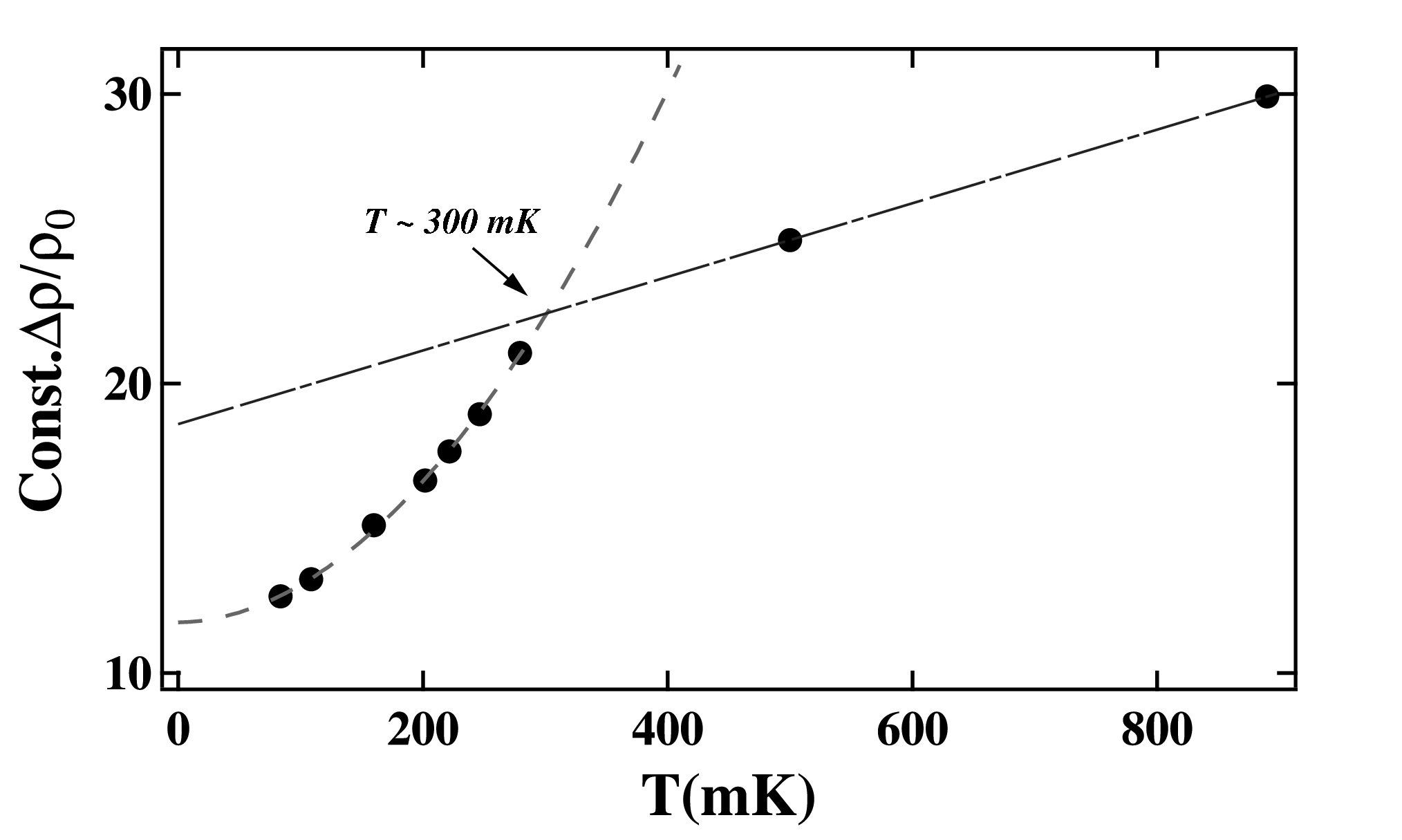}
\caption{\label{Resis} Change in behavior of the normal state resistivity with temperature 
at 12 T.  The change suggests FL to NFL behavior at the temperature is increased. The lines are  power law and linear fits. }
\end{center} \end{figure}

As we have mentioned, the large jump in penetration depth that we associate with the first order upper critical field matches very well with specific heat and data previously obtained by other techniques, as seen in Fig.~\ref{Phasedia}.
Yet, as can be seen in Fig.~\ref{rawdata}, with increasing temperature, the
height of the first order transition increases. This increase is due to the change
in resistivity of the normal state. By measuring the relative
change in frequency with temperature, at a field of $B=12$ T
parallel to the \emph{ab} planes, we were able to observe the relative
change in resistivity in the field-induced normal state (Fig.~\ref{Resis}). For temperatures below T $\approx$ 300 mK a
variable power law fit of the data yields a power of 1.92 which
is remarkably close to 2, the value expected for Fermi liquid (FL) behavior. In
contrast, the two points at higher temperature clearly depart from
the power law curve, indicating a change in the behavior of the
system to a non Fermi liquid (NFL). We are aware of very recent similar results obtained from
a direct measurement of the resistivity \cite{Paglione} and the Sommerfeld constant, $\gamma$ \cite{Bianchi2}, although at 12 T, Ref. \onlinecite{Paglione}  shows the transition to NFL behavior at a higher temperature than what we observed. (We note that other penetration depth studies have also found evidence for NFL behavior, but at zero magnetic field \cite{Ozcan, Chia}). At 5.5 T with the 
field perpendicular to the  \emph{ab} planes we only see a linear dependance of the resistivity indicating NFL behavior. This result is consistent with Refs. \onlinecite{Paglione} and \onlinecite{Bianchi2} where the crossover to FL behavior is observed above 7 T. 
This change in behavior leads us to believe that the same parameter may 
 be responsible for the suppression of the FFLO state at higher temperatures \emph{and} for the
change in resistivity behavior.  The coefficient F$^a_0$, which measures the interaction strength between  quasi-particles, could be this parameter. A larger positive value of $F^a_0$, and therefore a
stronger electron-electron interaction results in a T$^2$ variation
of resistivity with temperature, \cite{Gross} and at the same
time increases the range of stability of the FFLO state by lowering
the Pauli paramagnetic susceptibility of the Fermi
liquid system ($\chi_e \simeq \frac{\chi0_e}{(1+F^a_0)}$ where $\chi_e$ is the normal fluid susceptiblility) \cite{HBurk}.  According to Burkhardt \cite{HBurk} as  $F^a_0$ becomes smaller, the FFLO state is stable over a smaller temperature and field range and disappears for $F^a_0  < -0.5.$  It is interesting to note that  $(1+F^a_0)^{-1}$ =  R, Wilson's ratio.  We estimated R to be near 1 based on recent $\gamma$ and $\chi_e$ measurements, although the absolute value of R is unclear because of the problems of isolating the electron paramagnetic susceptibility, as mentioned above.  Nevertheless, below T = 2 K $\chi_e$ is constant and $\gamma$ increases and thus the trend is for R to decrease, which, consistent with Burkhart, stabilizes the FFLO state.

We would like to acknowledge R. MacKenzie and N.  Fortune for helpful discussions and A. Powell for building some of the electronics.  Funding for this project came from the NHMFL In House research and Visiting Scientist programs.  The NHMFL is supported by the NSF (Cooperative Agreement DMR-0084173) and the State of Florida.

\end{document}